\documentclass[iop]{emulateapj}
\usepackage{epsfig}
\usepackage{graphicx}
\usepackage{amsmath}
\usepackage{ifthen}
\usepackage{afterpage}
\usepackage{appendix}
\begin{document}
\shorttitle{CARRS I}
\shortauthors{Kollmeier et al.}
\def\feh{{\rm [Fe/H]}}
\newcommand{\msun}{M_{\odot}}
\newcommand{\kms}{\, {\rm km\, s}^{-1}}
\newcommand{\mas}{\, {\rm mas\, yr}^{-1}}
\newcommand{\cm}{\, {\rm cm}}
\newcommand{\gm}{\, {\rm g}}
\newcommand{\erg}{\, {\rm erg}}
\newcommand{\kel}{\, {\rm K}}
\newcommand{\kpc}{\, {\rm kpc}}
\newcommand{\mpc}{\, {\rm Mpc}}
\newcommand{\seg}{\, {\rm s}}
\newcommand{\kev}{\, {\rm keV}}
\newcommand{\hz}{\, {\rm Hz}}
\newcommand{\etal}{et al.\ }
\newcommand{\yr}{\, {\rm yr}}
\newcommand{\mpyr}{{\rm mas}\, {\rm yr}^{-1}}
\newcommand{\gyr}{\, {\rm Gyr}}
\newcommand{\eq}{eq.\ }
\newcommand{\statpi}{S{$\pi$}}
\def\var{{\rm var}}
\def\cc{{\rm cc}}
\def\ch{{\bf changed\ }}
\def\eff{{\rm eff}}
\def\arcsec{''\hskip-3pt .}
\newcommand{\bdv}[1]{\mbox{\boldmath$#1$}}
\title{\bf The Absolute Magnitude of RRc Variables From Statistical Parallax}
\author{
Juna A. Kollmeier \altaffilmark{1}, 
Dorota M. Szczygie\l\altaffilmark{2}, 
Christopher R. Burns\altaffilmark{1},  
Andrew Gould\altaffilmark{2}, 
Ian B. Thompson\altaffilmark{1}, 
George W. Preston\altaffilmark{1}, 
Christopher Sneden\altaffilmark{3},
Jeffrey D. Crane\altaffilmark{1}, 
Subo Dong\altaffilmark{4}, 
Barry F. Madore\altaffilmark{1}, 
Nidia Morrell\altaffilmark{1},
Jos{\'e} L. Prieto\altaffilmark{1,5}, 
Stephen Shectman\altaffilmark{1},  
Joshua D. Simon\altaffilmark{1}, 
Edward Villanueva\altaffilmark{1}}

\altaffiltext{1}{Observatories of the Carnegie Institution of Washington,
  813 Santa Barbara Street, Pasadena, CA 91101}
\altaffiltext{2}{Department of Astronomy, The Ohio State University,
  4051 McPherson Laboratory, Columbus, OH, 43210}
\altaffiltext{3}{Department of Astronomy, University of Texas at Austin}
\altaffiltext{4}{Institute for Advanced Study, 500 Einstein Drive, Princeton NJ}
\altaffiltext{5}{Department of Astrophysical Sciences, Princeton University, Princeton, NJ}

\begin{abstract}

We present the first definitive measurement of the absolute magnitude of RR
Lyrae c-type variable stars (RRc) determined purely from statistical
parallax.  We use a sample of 247 RRc selected from the All Sky
Automated Survey (ASAS) for which high-quality light curves, photometry and
proper motions are available.  We obtain high-resolution echelle
spectra for these objects to determine radial velocities and
abundances as part of the Carnegie RR Lyrae Survey (CARRS).  We find that $M_{V,\rm RRc} = 0.52 \pm 0.11$ at a mean metallicity of ${\rm [Fe/H]} = -1.59$.  This is to be compared with
previous estimates for RRab stars ($M_{V,\rm RRab} = 0.75 \pm 0.13$) and
the only {\it direct} measurement of an RRc absolute magnitude (RZ
Cephei, $M_{V, RRc} = 0.27\pm 0.17$). We find the bulk velocity of the halo to be $(W_\pi, W_\theta, W_z) = (10.9,34.9,7.2)\kms$ in the radial, rotational and vertical directions with dispersions $(\sigma_{W_\pi}, \sigma_{W_\theta}, \sigma_{W_z}) =  (154.7, 103.6, 93.8) \kms$.  For the disk, we find $(W_\pi, W_\theta, W_z) = (8.5, 213.2, -22.1)\kms$  with dispersions $(\sigma_{W_\pi}, \sigma_{W_\theta}, \sigma_{W_z}) =  (63.5, 49.6, 51.3) \kms$. Finally, we suggest that UCAC2 proper motion errors may be overestimated by about 25\%.

\end{abstract}

\section{Introduction}
Determining distances by use of multiple methods has a long and distinguished history in astronomy from antiquity to the present day.  Aristarchus of Samos first determined the Moon-Earth distance from the lunar eclipse.  A century later, Hipparchus checked Aristarchus' values using the independent method of terrestrial parallax: the position of the lunar limb during solar eclipse as seen from Alexandria and Hellespont.  More recently, astronomers have demanded that multiple methods be employed for zeroing in on the precise parameters governing the currently observed and mysterious accelerated expansion of the universe \citep{albrecht06}.

Pulsating variables have enjoyed a privileged role in the local volume.  Due to their characteristic light curves and relatively bright absolute magnitudes compared to the bulk of main sequence stars, they can easily be identified and readily measured in nearby galaxies.  With a local zeropoint for these systems, it is straightforward, modulo metallicity and reddening issues, to determine the distances to external systems.  Cepheids, and their more common, albeit fainter, relatives within the instability strip, the x (RRL) variables, have been two key elements in the historical endeavor to launch humanity out of the solar system and Milky Way and into the local cosmos.  Indeed the Cepheids currently serve as the anchor of the cosmological distance scale, having allowed the most precise measurement the local rate of expansion of the universe, $H_0$, to date (Freedman et al. 1994, Freedman et al. 2001).  In principle, once properly calibrated, all stars of known absolute magnitude should yield identical measurements of the distances to nearby galaxies.  This has not historically been the case. 

In particular, the RRL $M_V$ calibration has varied by almost half a magnitude depending on the method adopted, and, therefore, consistency with the Cepheid distance scale as well as other distance metrics has been difficult to establish.  As a result, attempts to determine RRL absolute magnitudes were largely abandoned over the past decade with a few notable exceptions (e.g., Dambis et al. 2009).   However, recently there have been new efforts.   Benedict et al. (2011) used HST trigonometric parallaxes of 5 RRL stars (including 4 RRab and 1 RRc variable) to obtain an average $M_V = 0.45 \pm 0.05$.  Klein et al. (2011) recently used mid-IR data from the WISE satellite to infer a mid-IR Period-Luminosity relation.

In this work, we present a third measurement of RRL absolute magnitudes using the method of Statistical Parallax (\statpi).   The large number of RRL that have been discovered in the last decade allow us to make a fresh assault on this issue.  Historically, RRL star distance measurements from  \statpi\ have come in systematically shorter than other distance indicators, in particular Cepheids (Barnes \& Hawley 1986, Hawley et al. 1986, Layden et al. 1996 (hereafter, L96),  Popowski \& Gould 1998a, Popowski \& Gould 1998b , Gould \& Popowski 1998 (hereafter, collectively PG$^3$), Dambis et al. 2009).  It is not yet fully understood why either this method or these two classes of objects should yield different distances to the same galaxies.   Thanks to automated synoptic all-sky surveys like the All Sky Automated Survey (ASAS, Pojmanski 2002), the number of RRL stars that have reliable light curves has increased by a factor $\sim$5 relative to the previous ``state-of-the-art".  The ASAS program has identified approximately 2000 RRL stars with 300-500 epochs of photometry.  By obtaining high-resolution spectra for these targets, we can both measure the radial velocities and metallicities needed for \statpi\ and address outstanding issues of systematics.  The light curves allow us to accurately determine pulsation phases and permit the measurement of the radial velocity at a single fiducial phase at which the pulsation velocity equals the star's systemic velocity (see Kollmeier et al. 2009).  Traditionally, obtaining the radial velocity component for RRL was laborious, requiring multiple epochs of spectroscopic observation.  The determination of the phase-velocity relationship for RRabc variables (Liu 1991, Kollmeier et al. 2009, Preston et al. 2011) allows a far more efficient strategy for obtaining critical radial velocity information with which to compute \statpi\ as we discuss further in Section 3.  

Historically, RRc variables have been either excluded from \statpi\ analyses or only approximately analyzed.  This is primarily due to two factors.  First, their hotter temperatures make it more challenging to determine abundances from low-resolution, low signal-to-noise ratio (SNR) spectra (see Layden 1994) and, as a result, these objects cannot be robustly classified by population (halo/disk) as required by modern \statpi.  However, high-resolution echelle observations circumvent this issue and allow, for the first time, a definitive \statpi\ analysis from RRc variables alone.  Second, there are fewer RRcs relative to RRabs, and it is only now that samples are large enough to perform a robust, self-consistent, pure RRc \statpi\ analysis.  We analyze our full (RRab + RRc) sample in a future work (Kollmeier et al. in preparation) and restrict our attention here to our RRc sample.

In Section 2 we present a brief overview of \statpi\ to remind the reader of the basic principles of the technique.  In Section 3 we present our sample selection, observations, data reduction, and analysis methods.  In Section 4 we review our updated methodology for determining \statpi, the results of which are discussed in Section 5 and compared to previous \statpi\ results in Section 6.  Finally, in Section 7 we discuss our results in light of recent and historical works on the absolute magnitude scale of RRL variables.

\section{Statistical Parallax}
The basic principle that underlies statistical parallax is that the absolute magnitude of any stellar population characterized by a particular velocity ellipsoid should have a single true value when derived from either transverse or radial kinematics of that tracer population.   The radial velocity (RV) determination of the ellipsoid is independent of any assumptions about distance, but the transverse determination requires an assumed value of the absolute magnitude.  The RVs alone yield values for the the velocity ellipsoid in units of $\kms$.  The proper motions, when scaled by the square root of the flux ($\mu_{scaled}=\mu \times 10^{V/5}$), yield values for the velocity ellipsoid in units of $\mas$.  The ratio of these two is a distance, D, which is the distance that an RRL star would have to be in order to have $V=0$.  And hence, the absolute magnitude of the RRL stars is $M_V= -5 {\rm log}(D/10 {\rm pc})$.

\subsection{Basic Formulae}

The basic formulae for computing statistical parallax in the presence of observational errors have been well-established (Clube \& Dawe 1980a,b; Murray 1983; Hawley et al. 1986; Strugnell, Reid, \& Murray 1986, PG$^3$).  The method involves a 10-parameter maximum likelihood fit to the kinematic and photometric data: the distance scale $\eta$, the
3 first moments (``bulk velocity'') $W_i$ of the sample velocity
distribution, and the 6 independent second
moments, $\sigma^2_{ij}$.  The key parameter of interest for distance scale determination is the value of the true versus assumed fiducial absolute magnitude of the tracer population.  
\begin{equation}
\eta = 10^{(M_{\rm fiducial} - M_{\rm true})/5}
\end{equation}
The error in determining this distance scaling is given by (Popowski \& Gould 1998a):
\begin{equation}
\frac{\sigma (\eta)}{\eta} =  \frac{1}{\sqrt N_\eff} \left( \frac{4}{3} + \frac{2 \kappa^2}{9} \right)^{-1/2}
\rightarrow 0.64\,N_\eff^{-1/2}
\label{eqn:spi_err}
\end{equation}
where 
\begin{equation}
\kappa = \sqrt{\sum_{ij} W_i \sigma^{-2}_{ij} W_j} \rightarrow W/\sigma
\end{equation}
and the last expression is for the (more intuitive) isotropic case
$\sigma_{ij} = \sigma \delta_{ij}$, $W$ is the bulk motion, $\sigma$ is the velocity dispersion, and
\begin{equation}
\frac{1}{N_\eff} \equiv \frac{2}{3}\left(\frac {1}{N_{RV}}  + \frac{1}{2N_{pm}}\right)
\end{equation}
where $N_{RV}$ is the number of stars with accurate RV measurements and $N_{pm}$ is the number of stars with accurate proper motion measurements. The evaluation in Equation~(\ref{eqn:spi_err}) is for $\kappa=2.2$, which is typical of halo RRL samples.

\subsection{Observational Requirements}
The distance scale accuracy improves as $N_\eff^{-1/2}$, i.e., the sample of stars with {\it accurate} proper motions and radial velocities is increased in size.  ASAS has provided a public catalog of RRL stars identified via high-cadence light curve analysis \citep{dorota07, dorota09}.  In addition to photometry from the ASAS survey, these stars have proper motions from the second USNO CCD Astrograph Catalog (UCAC-2; Zacharias et al. 2004), which covers the entire southern hemisphere.  Radial velocity information is not published for the majority of the Southern Sky and what is therefore required are spectroscopically determined velocities for this large sample of objects that has corresponding proper motions and photometry.  Before beginning our extensive observational program, we evaluated the suitability of the ASAS sample for measuring \statpi\ considering known systematic uncertainties.  We discuss each of these below.  

\subsubsection{ Multiple Populations}
The key underlying assumption of \statpi\ is that the RRL
are a faithful tracer population of a single velocity ellipsoid and do
not exhibit poorly mixed kinematics, for example, from coherent
stellar streams.  Clearly, as one goes to very large distance
(i.e., the outer halo) this assumption breaks down as RRL populations
become increasingly contaminated by satellite debris on non-mixed orbits.  
Indeed, this consideration alone prevents us from extending the survey to extremely faint
magnitudes where the distances probed preferentially contain such
kinematics (e.g. Kollmeier et al. 2009).  The distribution of distances for the ASAS RRc sample is shown in Figure~\ref{fig:distances}.    
As can be seen, the majority of our sample lies within a distance of 4$\kpc$.  The solar neighborhood on these scales is known to be sufficiently smooth that our statistics are unaffected by kinematic substructure (e.g., Gould 2003).

\subsubsection{UCAC-2 Proper Motions}
\begin{figure}[htb]
\centerline{
\includegraphics[width=3.5in]{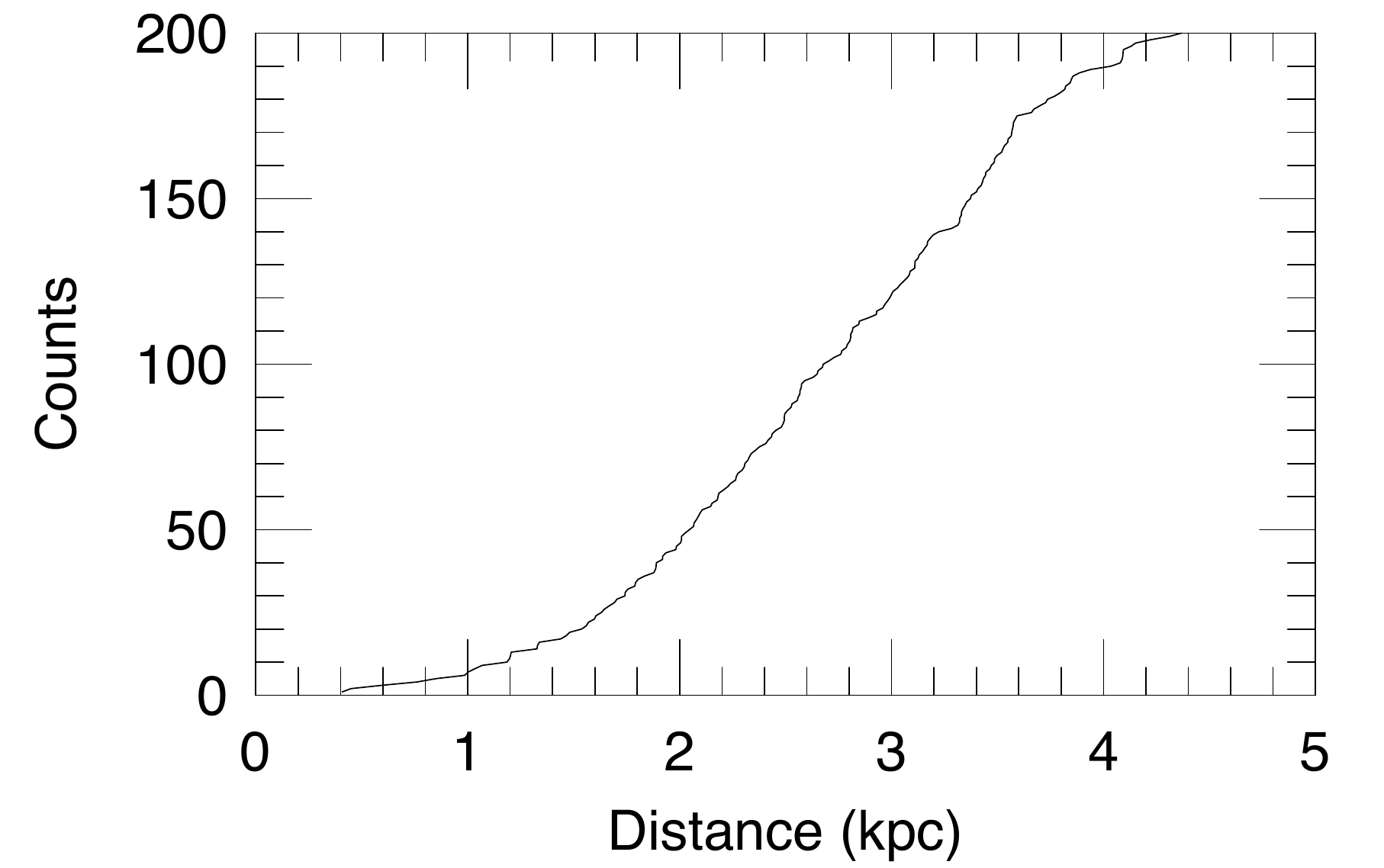}
}
\caption{Distribution of distances for sample stars.}
\label{fig:distances}
\end{figure}

The proper motion errors from UCAC-2 are typically 5-6 $\mas$ for the majority of the sample corresponding to $\sim(26 \kms) \times ({\rm D/kpc})$.    At first sight, the proper motion errors may seem too large to be useful, given that the median distance is about $D_{\rm med}=2.7\,$kpc (see Fig.~\ref{fig:distances}, determined using an initial value for the absolute magnitude scale of $M_V=0.5$),  and the direction-averaged dispersion is about $\sigma\sim 120\,\kms$. Together, these imply that the errors are typically a large fraction of the quantity being measured.  In fact, \statpi\ is extremely robust with respect to
observational errors.  Equation (17) of Popowski \& Gould (1998a) (hereafter PG98a) shows that the statistical error should increase by {\it only a few percent} relative to \rm Equation~(\ref{eqn:spi_err}), given these errors.  We verify that this is actually the case in Section~\ref{sec:results}.  Equation (18) of PG98a shows that one must be careful about systematic errors that can result from {\it misestimating} the size of these errors.  Indeed, the \statpi\ methodology contains an internal check on the proper motion errors.  

\subsubsection{Reddening \& Photometry}
Since \statpi\ uses apparent magnitudes, it is important to have robust extinction measurements for objects in the sample.  We use the reddenings provided by Schlegel, Finkbeiner \& Davis (SFD, 1998) and a standard extinction coefficient, $R_V=3.1$ to determine $A_V$ \footnote{We note that updates to the SFD extinction maps have been provided by Schlafly \& Finkbeiner (2012) and Peek \& Graves (2011).  These new maps qualitatively agree, sharing the conclusion that the SFD extinction map is generally thought to be overestimated, but do not quantitatively agree in terms of the precise magnitude of this difference (14\% and 2\% in each study).  The effect of adopting the more extreme correction leads to a correction in $\eta$ of $\Delta \eta = 0.03 \times log(10)/5 = 0.014$.  We adopt the SFD values until a definitive calibration is agreed upon.}. Because many of our sources are at high Galactic latitude (see Figure~\ref{fig:skydist}), this directly gives adequate values for these sources.  We adjust the reported reddening values to account for the finite distance from the plane of our objects.  In some instances, our objects are too close on the sky with other field stars in projection to yield reliable ASAS photometry (see Pojmanski (2002) for details of ASAS photometry).  These objects are eliminated from our analysis.

With secure photometry and transverse kinematic measurements in hand, we began our spectroscopic survey.  

\section{Observations \& Data Reduction}
\medskip

\begin{figure}[htb]
\centerline{
\includegraphics[width=3.5in]{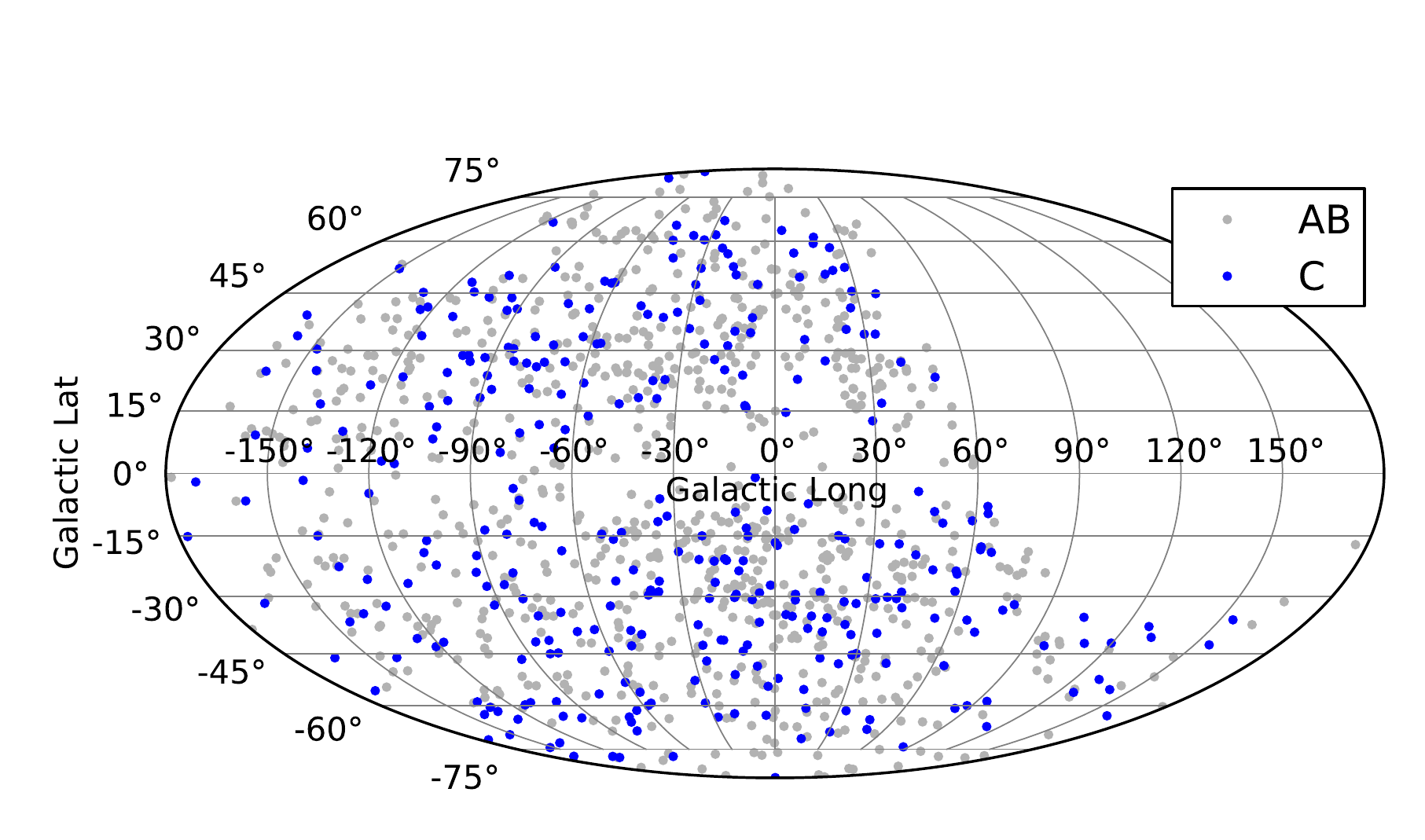}
}
\caption{Sky distribution of our ASAS targets.  The RRab sample is shown in grey points and the RRc sample sample is shown in blue.  The analysis presented in this paper includes 247 of the RRc sample points.}
\label{fig:skydist}
\end{figure}

The observations presented here were made in 2011 and 2012 with the echelle spectrograph mounted on the 2.5m du Pont telescope at Las Campanas Observatory as part of the larger Carnegie RRL Survey (CARRS; Kollmeier et al. in preparation).  Upon completion, our survey will have moderate signal-to-noise ratio spectra for approximately 1200 RRL stars observable from the Southern Hemisphere.  The sky distribution of our full survey sample is shown in Figure~\ref{fig:skydist}.  The grey points show RRab stars and the blue points show RRc stars.  Observations were designed to reach SNR$\sim 15$ in the order containing the Mg I triplet at 5170\AA\ at the target phase.  Exposures for a given target bracketed observations of a ThAr lamp through a 1.5x4-arcsec slit.

\subsection{Synoptic Observations}
\begin{figure}[htb]
\centerline{
\includegraphics[width=3.5in]{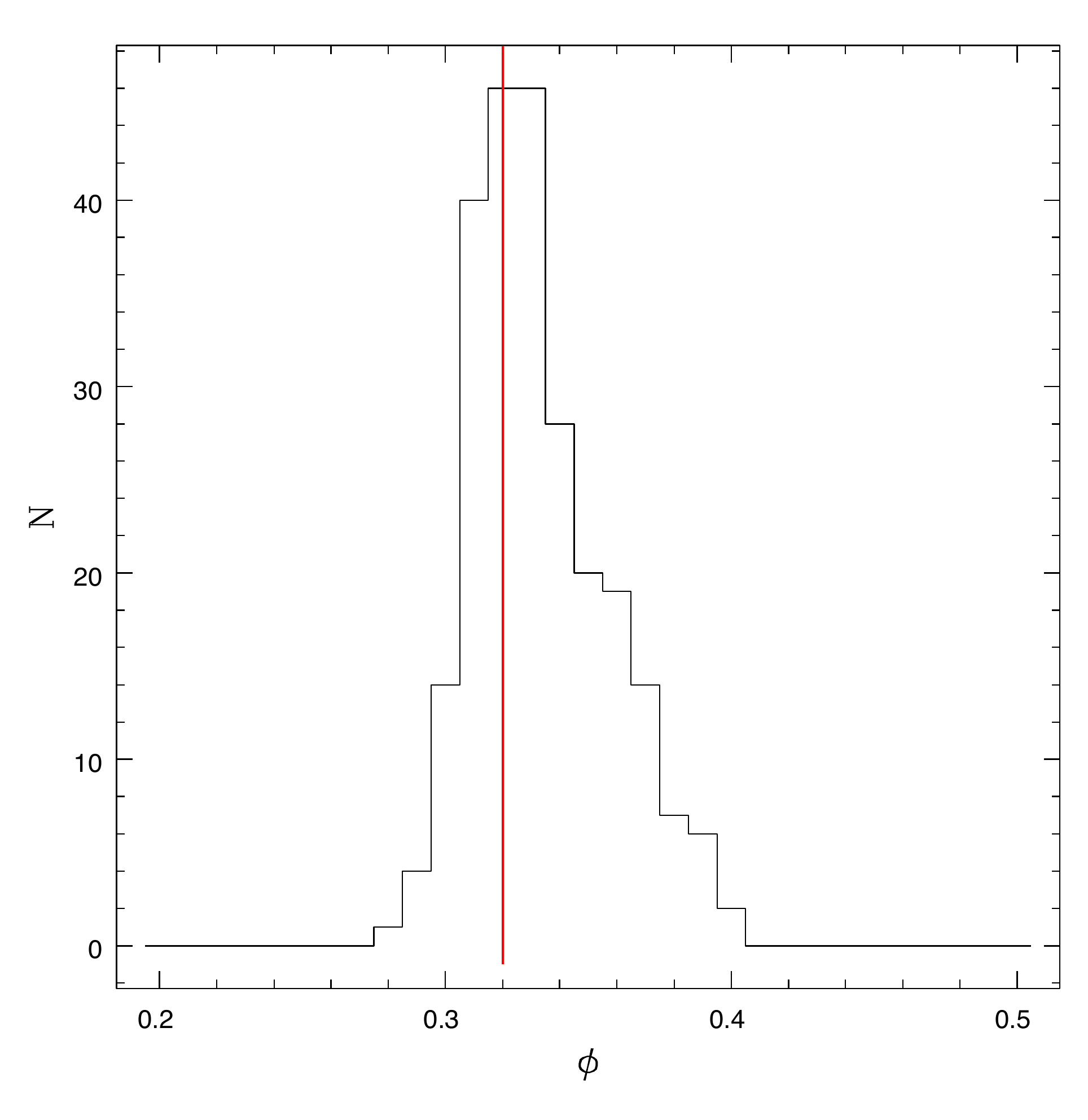}
}
\caption{Phase distribution of our sample of 247 RRc variables.  The red line in the figure shows the phase targeted for the RRc sample throughout our survey.}
\label{fig:phases}
\end{figure}

The photospheric velocity of RRL stars changes by many $\kms$ over the course of a pulsation cycle.  Traditionally, one has taken multiple observations through the pulsation cycle and fit the resultant velocity function (when the phase is known) in order to obtain the systemic velocity of the target (e.g. Layden 1994).  We adopt the time-averaged velocity of the pulsation curve as the center-of-mass velocity of the star as detailed in \citet{liu91} and \citet{preston11} for RRabs.  Integration of detailed velocity curves of the RRc variables T~Sex, TV~Boo, DH~Peg, and YZ~Cap, all measured with the du Pont echelle, shows that the pulsation velocity is equal to the star's time-average velocity at phase 0.32, reckoned relative to maximum light. Our observations were all made as close to this phase as possible, and velocity corrections were applied adopting the following correction:
\begin{equation}
\Delta \phi_{RRc} =  62.0 \kms (0.32 - \phi_{obs}).
\end{equation}
Figure~\ref{fig:phases} shows the resultant phase distribution for our targets.  Note that these velocity corrections are no larger than $5\kms$ for any given star.

\medskip
\subsection{Data Reduction Pipeline}
In order to ensure uniform quality across our sizeable database, a data reduction pipeline was constructed following Kelson (2003).  The pipeline divides the reduction process into first processing the calibration frames, and then using these to reduce the science frames and extract spectra.  The calibration stages involve processing the frames for flat-fielding, determining the order edges and y-distortion using sky flats, and obtaining wavelength solutions using a Th-Ar lamp frame.  Once these frames are produced, the science spectra are reduced.  The science spectra are first divided by the processed and normalized flat-field and then wavelength calibrated.  The spectra are then sky-subtracted and the spatial profiles of the resultant 2-D spectra are used to find extraction apertures.  The pipeline takes advantage of parallel processing, which speeds up the reduction process significantly for a large number of objects.

\medskip

Post-extraction processing of the spectra was done with the IRAF\footnote{ IRAF is distributed by the National Optical Astronomy Observatories, which are operated by the Association of Universities for Research in Astronomy, Inc., under a  cooperative agreement with the NSF.} {\sc ECHELLE} package.  Radial velocities were measured relative to a high SNR template of the metal-poor star HD140283 using the IRAF FXCOR routine.  The template spectrum was taken with the du Pont Echelle spectrograph in the same configuration as survey stars.  The heliocentric radial velocity for HD140283 has been determined, independently in four separate high-resolution studies, to be  $V_{\rm helio} = -170.92 \pm 0.19 \kms$ (e.g. Tsangarides et al. 2004, Aoki et al. 2002, Lucatello et al. 2005, Latham et al. 2002).   The cross
correlations were made on the wavelength interval 4900\AA\ -- 5500\AA.  Objects that were originally classified as RRL stars based on their light curves but, upon inspection of their spectra, were found to be binaries or other non-RRL variables were removed from the sample.

The radial velocity error estimates returned by fxcor are extremely small, so we estimate the true errors
by making repeat measurements of the {\it corrected} velocities
on a subset of 11 stars.   From these we determine
$\sigma_{\rm RV} = 1.96\kms$.  We add this in quadrature to
our mostly tiny formal errors.  We note that, from PG$3$, the
velocity errors enter the final result as
$\Delta\eta/\eta\sim (\sigma_{\rm RV}/\sigma_{\rm halo})^2/6 < 10^{-4}$
and so have no practical effect in any case.  We nevertheless include
this correction for completeness. 

\medskip
\subsection{Abundance Determination}
The measurement of absolute abundances for large samples of RRc stars has not been undertaken previously.  We therefore must calibrate our pipeline abundance measurements (which yield consistent relative abundance determinations) to the scant data currently available.  Based on analysis of the high SNR du Pont echelle spectrum of RRc variable YZ Cap, we adopted a single set of atmospheric parameters ($T_\eff=$7000K, $\log g=2.2$, $v_{\rm micro}=2.5\kms$, [$\alpha$/Fe]=+0.35) for all survey stars.  The value of $T_\eff$ is somewhat constrained by the small area occupied by RRc stars in the color-magnitude diagram, and our survey spectra are not of sufficiently high SNR to determine this parameter independently.  The abundance measurements in RRL stars are relatively insensitive to the adopted values for $\log g$ and $v_{\rm micro}$.  For our final calibration, we rely primarily on a detailed study (Govea et al. in preparation) that examines the sensitivity of RRc abundance determinations with respect to stellar parameters using a small sample of high SNR spectra also obtained with the du Pont echelle.  This is the only study we know of in which the sensitivity of RRc abundance measurements are systematically studied.  We also compare to the known distribution of RRab metallicities, as a secondary calibration.  We defer a more detailed analysis of this metallicity calibration to a future work where more high SNR, high-resolution data are available for comparison (Kollmeier et al. in preparation).

\begin{figure}[htb]
\centerline{
\includegraphics[width=3.5in]{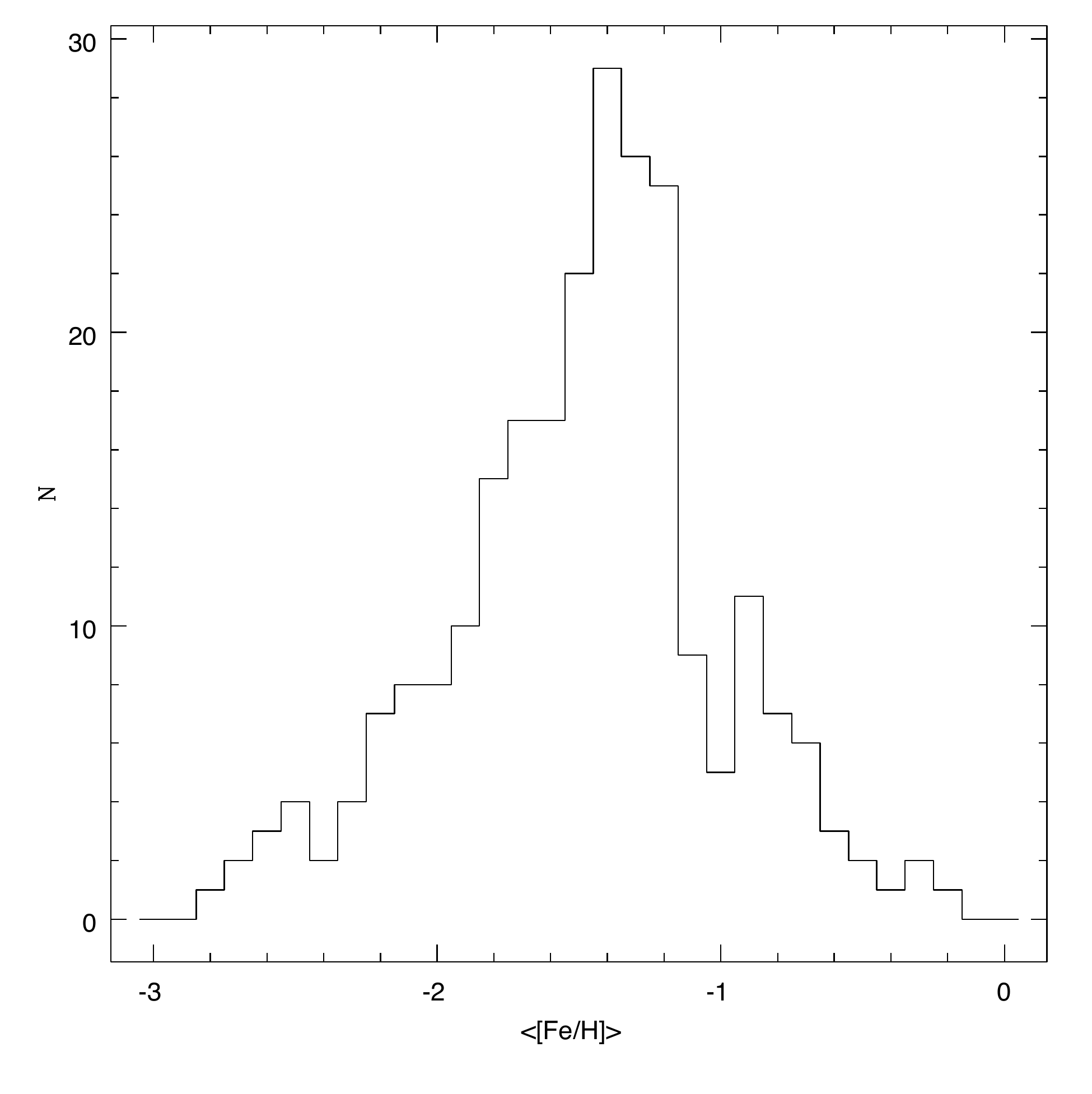}
}
\caption{[Fe/H] determinations for our sample of 247 RRc variables.  Abundances were determined by averaging two broad spectral windows which had maximal signal to noise as well as numerous spectral features. }
\label{fig:abundances}
\end{figure}

Pipeline abundance measurements were performed by fitting to a grid of synthetic spectra generated by MOOG\footnote{MOOG is a publicly available code to determine abundances in stars through LTE analysis.  MOOG is available at http://www.utexas.edu/~chris/moog.html}.  Spectra reduced by our spectroscopic pipeline were cleaned of remaining noise spikes and smoothed using a 3-pixel boxcar filter.  Owing to the relatively broad lines in RRL stars, this smoothing has essentially no effect on the abundance determination. RRL stars are warmer than the Sun and generally metal-poor and therefore contain few spectral features beyond 5500\AA.  As our SNR is maximized  between approximately 4400\AA\ and 5500\AA, we therefore used two broad spectral ranges for our synthetic fits:  4400-4680\AA\ and 5150-5450\AA .  The latter region contains primarily neutral-species transitions as well as the Mg I b lines.  The former region has numerous singly-ionized species including Ba II 4554\AA .  The synthetic spectral grid covers a range of effective temperature ($\rm T_\eff$), surface gravity ($\log g$), metal abundance and microturbulent velocity ($v_{\rm micro}$).  The atomic line lists for the syntheses were begun with the Kurucz line database, then refined until a good match was achieved for the spectrum of the Sun.  The smoothed synthetic spectra were compared with the observed spectra and the parameters of those with lowest $\chi^2$ were chosen as fits.  In Figure~\ref{fig:abundances} we show a histogram of our pipeline-derived and calibrated  abundances averaging the (independently) derived abundances for the two broad spectral windows.

\section{Disk/Halo Separation}
\begin{figure}[htb]
\centerline{
\includegraphics[width=3.5in]{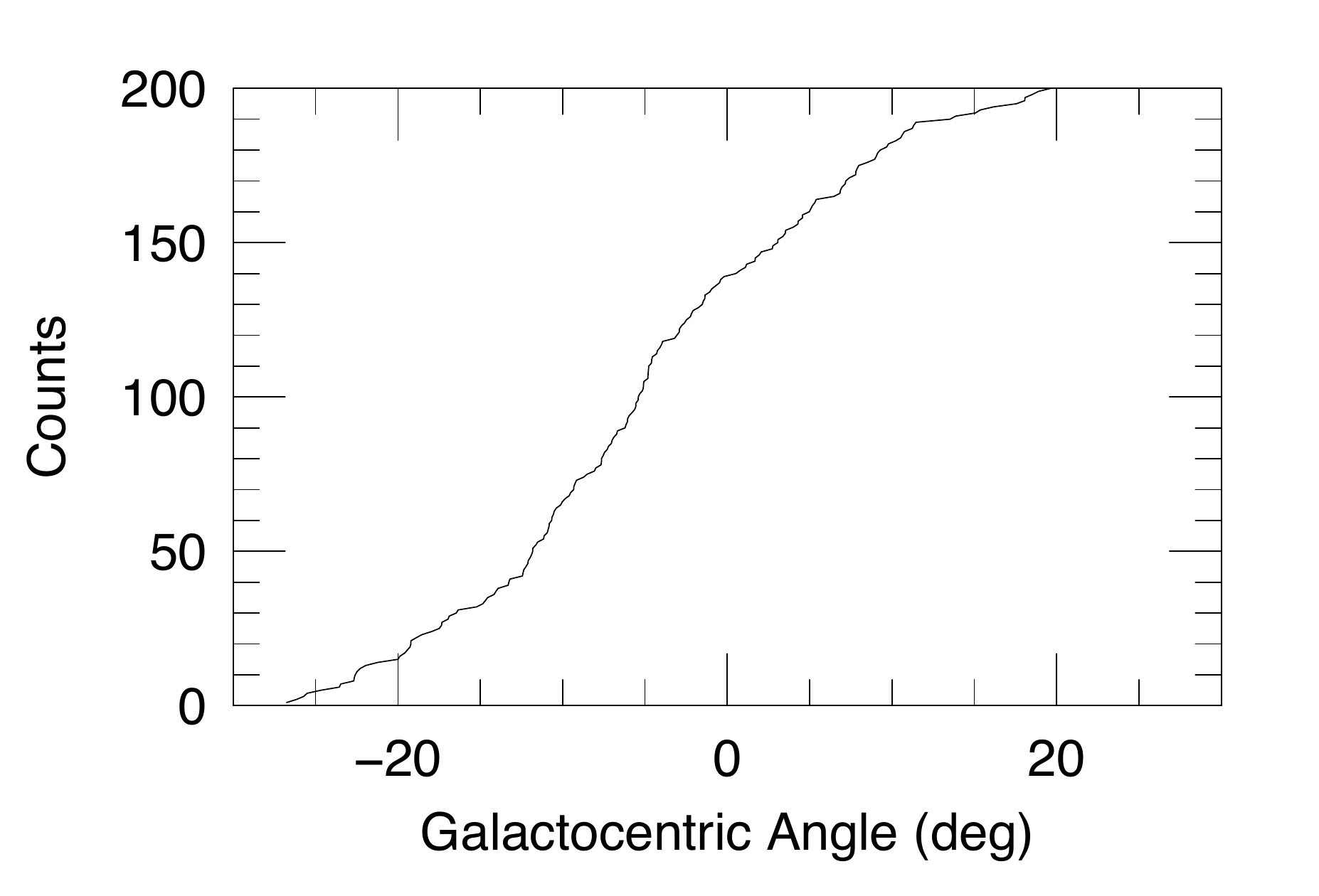}
}
\caption{Distribution of rotation angles between the Sun-centric and star-centric frames of reference.}
\label{fig:angles}
\end{figure}

\begin{figure}[htb]
\centerline{
\includegraphics[width=3.5in]{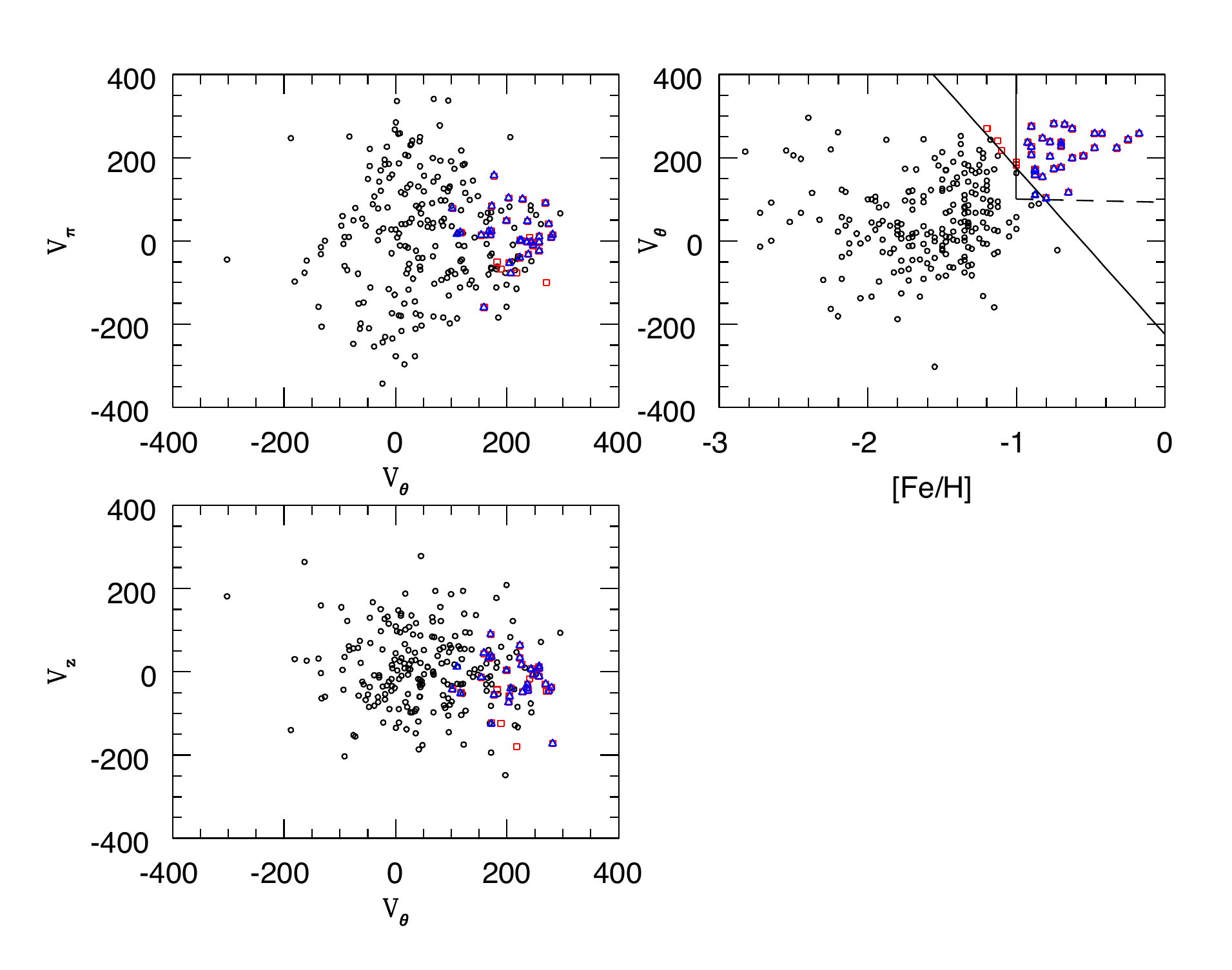}
}
\caption{Kinematics and abundance information for our 247 RRc stars.  The left panels show the values of $v_\pi, v_\theta,$ and $v_z$, that is, the velocities in the radial direction, the direction of Galactic rotation, and toward the Galactic pole respectively.  These velocities are corrected for rotation w.r.t. $U, V, W$ velocities.  The right panel shows the rotational velocity component as a function of [Fe/H].  In each panel, red squares denote the ``DISK-1" population and blue triangles show the ``DISK-2" population.  These stars are eliminated from our ML analysis as described in the text.  }
\label{fig:disk_halo}
\end{figure}

With the kinematic and chemical information measured as described above, we are prepared to segregate our sample into the Halo and Disk populations.  We convert our proper motions and radial velocities to radial, rotational, and vertical velocities ($U,V,W$) as per L96.  We correct these velocities for the dynamical solar motion (+9, +250, +7 $\kms$) and we rotate the Sun-centric ($U,V,W$) 3-space velocity to the local frame of the star assuming cylindrical symmetry of the Galaxy.  The cumulative distribution of these rotation angles is shown in Figure~\ref{fig:angles}.  Note that they span the range of roughly $\pm 25^\circ$. The resultant velocities $(v_\pi, v_{\theta}, v_{z})$ for our sample are shown in the left panels of Figure~\ref{fig:disk_halo} for direct comparison with Figure 3 of L96 which shows the same quantities for their sample of RRab stars.  Similarly, the right hand panel of Figure~\ref{fig:disk_halo} shows the rotational velocity component as a function of measured abundance for direct comparison with Figure 4 of L96.  It is important to remind the reader again that we are using RRc stars rather than RRab stars as tracers of the velocity ellipsoid, in contrast to L96.  Despite this difference in tracer population, the kinematics of this sample are similar to those of L96.

While it is not possible to know with certainty whether any particular star is a member of the disk or halo,  we can clearly see a track of stars that has chemical and kinematic properties similar to a disk-like population.  These stars are relatively high metallicity and are rotating in the direction of the Sun with similar speed.  The exact demarcation of the disk/halo line is necessarily uncertain.   Following L96, we therefore perform our analysis on subsamples with the aim of determining how sensitive our results are to these uncertainties.  As we will show, they are not.  We identify two halo/disk subsamples.  The first (HALO-1/DISK-1) considers all stars rightward (upward) of the line $v_\theta = 400 {\rm [Fe/H]} - 225$ to be members of the disk/thick disk and they are excluded from the halo population of interest.  Of our 247 objects, this procedure assigns 34 to the disk and 213 to the halo.  These objects are shown by the red squares in Figure~\ref{fig:disk_halo}.  The second (HALO-2/DISK-2) considers all stars with [Fe/H]$>-1.0$ {\it and} $v_\theta > 100 \kms$ to be disk stars.  Objects satisfying these criteria are shown as blue triangles in Figure~\ref{fig:disk_halo}.  Of our 247 objects, this classification eliminates 30 leaving 217 for our \statpi\ analysis.  We note that our HALO-1/DISK-1 and HALO-2/DISK-2 designation is slightly modified from L96 given our sample.   However, our results are relatively insensitive to these distinctions, although our errors will necessarily scale with the square root of the number of retained objects.

\medskip
\section{Results}\label{sec:results}
\subsection{Halo Velocity Ellipsoid}
We carry out a 10-parameter Markov Chain Monte Carlo (MCMC) 
maximum-likelihood fit to the data.  These include the nine kinematic parameters describing the velocity ellipsoid (three diagonal elements corresponding to the bulk motions and 6 dispersions) and one parameter describing the absolute magnitude scaling $\eta$.  Table~\ref{tab:results} gives the values and errors of the 10 \statpi\ parameters for our samples.  For reasons explained in Appendix A, their errors
are almost perfectly Gaussian.  Therefore the 10-dimensional likelihood surface is completely specified by these numbers, together with the correlation coefficients, which are also given in Appendix A. The fact that the parameters derived from these two different fits are consistent at well below $1\,\sigma$ shows that our choice of disk/halo separation does not significantly affect our results.

\begin{deluxetable*}{c r r r r r r r r r r}
\tablecaption{Resultant Parameters and Errors.\label{tab:results}}
\tablehead{
\colhead{Sample} &
\colhead{$\eta$} &
\colhead{$w_1$} &
\colhead{$w_2$}&
\colhead{$w_3$}&
\colhead{$C_{11}^{1/2}$}&
\colhead{$C_{22}^{1/2}$}&
\colhead{$C_{33}^{1/2}$}&
\colhead{$\tilde{C}_{12}$}&
\colhead{$\tilde{C}_{13}$}&
\colhead{$\tilde{C}_{23}$}\\
\colhead{ } &
\colhead{($\sigma_\eta$)} &
\colhead{($\sigma_{w_1}$)} &
\colhead{($\sigma_{w_2}$)}&
\colhead{($\sigma_{w_3}$)}&
\colhead{($\sigma_{C_{11}^{1/2}}$)}&
\colhead{($\sigma_{C_{22}^{1/2}}$)}&
\colhead{($\sigma_{C_{33}^{1/2}}$)}&
\colhead{($\sigma_{\tilde{C}_{12}}$)}&
\colhead{($\sigma_{\tilde{C}_{13}}$)}&
\colhead{($\sigma_{\tilde{C}_{23}}$)}\\
}
\startdata

HALO-1 &  1.007  & 11.8 &  32.5 &  8.3  & 155.6 & 102.4 & 93.8 &  0.037  & -0.006 &  -0.114 \\
 - &  0.048 & 11.1 &  9.9  &  6.9 &  10.2 &  6.0  &  5.6  &  0.075  &  0.075   & 0.076 \\
{\tiny Analytic-err} & 0.045 &  10.7 &   9.6 &   6.4  & 8.9  &  5.8 &  5.3 &  0.069  & 0.069 &   0.069 \\
HALO-2 &  1.002  & 10.1 &  37.4 &   6.1  & 153.8 & 104.8 &  93.8  & 0.018 &   0.008 & -0.144 \\
 - &   0.048 & 10.9 &   10.0 &  6.8 & 10.1 &  6.1 &  5.6 &   0.074  & 0.075 &  0.075\\
DISK-1 & 0.728  & 4.0 & 217.1 & -24.9 &  62.1 & 42.6 & 49.4  & 0.000  & -0.192 &   0.008 \\
  - &  0.142 & 11.3 &  9.5 & 10.8 & 10.2  &  8.7 & 12.8  &  0.198 &  0.193  & 0.207 \\
 DISK-2 & 0.796 &13.1 & 209.3 & -19.2 & 64.9 & 56.5 &  53.1 &  -0.020  & -0.397 & -0.088\\
- &   0.158 & 12.6 & 12.9 & 11.7 & 11.2 & 12.2 & 14.1  & 0.203  & 0.190   & 0.207\\
  \enddata
\end{deluxetable*}

For our HALO-1 sample, we obtain a value of the absolute magnitude of RRcs of $M_V = 0.47\pm 0.10$ and for our HALO-2 sample, this value is $M_V = 0.49 \pm 0.10$.  We discuss these values in the context of previous measurements in Section~\ref{sec:comp}.

{\subsection{Test of Accuracy of UCAC-2 Proper-Motion Error Estimates}\label{sec:ucacpm}}
In the analysis reported above, we adopted the proper-motion error estimates
in the UCAC-2 catalog.  As discussed by PG98a,
if the size of these error estimates were systematically too big, it would
induce a systematic error in our distance-scale estimate, making the RRL
appear to be systematically further (hence more luminous) than they actually
are.  This is a potential concern because Zacharias et al. (2004)
report external tests only on their mean proper motion estimates and not 
on their error estimates.  Fortunately, we are able to perform an internal
test on these error estimates.  This test shows first that the UCAC-2
error estimates are probably overestimated by about 25\%, and second
that this level of mis-estimation has almost no effect on our results.  We note that our RV errors are too small to have
any effect, whether correctly estimated or not.  They will therefore
be ignored in the following discussion.

We first review the basic physics that permit such a test and then
present results. What is now called ``statistical parallax'' was
formerly (in the first half of the last century) 
divided into two effects.  One, called ``secular parallax'',
compared the mean bulk motion ($W_i$) 
in RV with that in (flux-scaled) proper motions.
The other, also called ``statistical parallax'' compared the amplitude
of the dispersions ($\sigma_{ij}$) in RV and proper motions.  In modern \statpi,
these are done simultaneously in a single fit.  However, the errors in
the proper-motion error estimates enter very differently into these
two components.  For the $\sigma_{ij}$ comparison an overestimate of the
errors leads to a corresponding underestimate of the intrinsic proper
motions of the stars, which can be reconciled with the dispersions measured
in RV only by placing them further away.  However, for the $W_i$ measurement,
which is first-order (not second-order) in the proper motions, there
is no such effect.  If one wished to be completely safe from any such
error-bar misestimates, one could in principle choose to just compute
the ``secular parallax''.

However, the maximum likelihood formulation of \statpi\ permits a more
sophisticated approach.  The likelihood function is a direct and
sensitive indicator of the consistency of the two components of the
\statpi\ determination when the error bars are systematically rescaled. Table~\ref{tab:rescale} shows results for several such rescalings.  The log-likelihood
is maximized for a rescaling factor $f_{\rm err}=0.5$, while $f_{\rm err}=1$
yields a log-likelihood that is lower than this $\Delta\log L=3.8$.
The probability of this occurring by chance is about 
$\exp(-\Delta\log L)\sim 2\%$.  While this is certainly not an impossible 
statistical fluctuation, neither is there any compelling prior that the
UCAC-2 errors are exactly as given.  We note that at $f_{\rm err}=0.8$, 
$\Delta\log L=1.5$, which is very plausibly due to statistical fluctuations.
Because we do have some prior that these errors are not grossly overestimated,
and because there is no statistical evidence that $f_{\rm err}<0.8$, we
adopt this as our best estimate of the errors.  We note from Table 1
that the result of changing $f_{\rm err}$ from 1.0 to 0.8 is to reduce the
distance scale by approximately 1\%, which is about 20\% of our statistical
error.  The remaining parameter estimates barely change as well.

\begin{deluxetable*}{c r r r r r r r r r r r}
\tablecaption{Effect of Proper Motion Error Rescaling \label{tab:rescale}}
\tablehead{
\colhead{Scale Factor}&
\colhead{$\Delta log \ell$}&
\colhead{$\eta$} &
\colhead{$w_1$} &
\colhead{$w_2$}&
\colhead{$w_3$}&
\colhead{$C_{11}^{1/2}$}&
\colhead{$C_{22}^{1/2}$}&
\colhead{$C_{33}^{1/2}$}&
\colhead{$\tilde{C}_{12}$}&
\colhead{$\tilde{C}_{13}$}&
\colhead{$\tilde{C}_{23}$}\\
}
\startdata
1.0 &  3.91& 1.007 & 11.8 &  32.5 &   8.3 & 155.6 & 102.4 &  93.8 &   0.037 &  -0.006 &  -0.114 \\
0.8 & 1.51 & 0.994 & 12.0 &  34.1 &   8.2 & 155.8 &  102.9 &  94.4 &   0.033 &   -0.003  & -0.106 \\
0.7 & 0.74 &  0.987 & 12.1 &  35.0 &   8.2 &  155.8 & 103.1 &  94.8 &   0.031 &  -0.002&   -0.102 \\
0.6 & 0.30 & 0.981 & 12.2 &  35.7  &  8.2 & 156.0 & 103.3 &  95.4 &   0.029 &  -0.001 &  -0.098 \\
0.5 & 0.09 & 0.974 & 12.4 & 36.6  &  8.1 &  155.9 & 103.4 &  95.8 &   0.028   & 0.000&   -0.094 \\

\enddata
\end{deluxetable*}

\subsection{Small Correction for Malmquist Bias}

Because the sample is magnitude limited and the target
population
has an intrinsic dispersion in absolute magnitude, $\sigma_{M_V}$, the
sample contains more stars with brighter-than-average luminosity
than fainter-than-average.  This leads to a correction
\begin{equation}
\Delta \eta/\eta = -3\biggl({\sigma_{M_V}\over 5/\ln 10}\biggr)^2
\label{eqn:malm}
\end{equation}
We modify the PG98a estimate to obtain $\sigma_{M_V}=0.1$ as follows.
PG98a began with the observed dispersion in the Large Magellanic Cloud (LMC) 
of 0.17 mag.  They estimated that 0.09 mag was due geometric dispersion, and
so estimated an intrinsic dispersion of 0.14 mag.  We further note that for an estimated metallicity
dispersion of 0.5 mag, and assuming a slope of $K=0.214$ 
mag dex$^{-1}$ in the
RRL $M_V$-[Fe/H] relation, an additional 0.1 mag of scatter can be 
accounted for by metallicity variation.  Subtracting this in quadrature,
we obtain 0.1 mag intrinsic dispersion.
This leads to a correction $\Delta\eta / \eta = -0.006$.

Note that PG98a identified another effect due to the dispersion of absolute
magnitudes, which 
scales $\Delta \eta/\eta \sim 0.25\sigma_{M_V}^2\sim 0.003$.
We briefly describe this effect and why it is proper {\it not} to include
it here.  If stars do not have exactly the same absolute magnitude,
then their mean square scaled (by flux) proper motions will be greater than
their squared scaled (by mean flux) proper motions.  This will mimic
a larger proper-motion dispersion and so cause the stars to appear
closer (so dimmer) than they actually are.  However, because the mechanism
of this effect is similar to that caused by misestimation of
the proper-motion errors (Section \ref{sec:ucacpm}), this small correction
is already absorbed into the correction adopted there.

\subsection{Final Result}

To obtain our final result, we incorporate the small corrections
for Malmquist bias and for probable overestimation of the UCAC-2
proper motion errors and take the average between our two halo samples,
\begin{equation}
M_V = 0.522 \pm 0.106\pm 0.031\quad {\rm  at} \quad 
\langle {\rm [Fe/H]} \rangle =-1.59
\label{eqn:mvfinal}
\end{equation}
The systematic error (0.031 mag) is determined by combining in quadrature
the variance from different definitions of the halo (0.010 mag), uncertainty
in the degree of overestimation of the proper motion errors (0.025 mag), and uncertainty in the extinction scale (0.014 mag).
There is also a systematic error of about 0.05 dex in the metallicity
scale, and so the mean metallicity at which this estimate is valid.

Because the statistical errors are about 3.5 times larger than the systematic
errors, a detailed investigation of the latter is not warranted at the
present time.  This issue will become somewhat more pressing when we 
combine the analysis of CARRS RRab and RRc stars.

Table 1 shows a comparison between the errors derived from the the MCMC
and those predicted analytically using the formulae of PG98a, which
assume zero measurement errors and isotropic sky coverage.  The actual
errors are only slightly larger than the ideal errors despite the
relatively large proper motion errors.  In Appendix A, we derive the
corresponding analytic estimates of the correlation coefficients among
the 10 \statpi\ parameters and show that these are in good agreement
with the MCMC determinations. 

\subsection{Disk Kinematics}
We can run our maximum likelihood machinery just as easily on a pure disk population as we can on a pure halo population, albeit with the substantially reduced numbers in our disk sample.  We perform this analysis and include the results in Table~\ref{tab:results}.   Interestingly, the value we obtain for absolute magnitude  (averaged between the two ``disk'' definitions) is $M_V = 1.08 \pm 0.43$ at [Fe/H]$=-0.7$. After taking account of the slope 
$K=0.214\,{\rm mag\,dex^{-1}}$ in the $M_V$-[Fe/H] relation, this
is consistent at the $1\,\sigma$ level with our result for the halo sample.  Also of interest are the detailed values of the velocity ellipsoid for the disk population: $(W_\pi, W_\theta, W_z) = (8.5, 213.2, -22.1)\kms$  with dispersions $(\sigma_{W_\pi}, \sigma_{W_\theta}, \sigma_{W_z}) =  (63.5, 49.6, 51.3) \kms$.  Indeed, it is thick-disklike, which gives us confidence that we are in fact removing a kinematically distinct population via our chemodynamical criteria.

\section{Comparison with Previous Parallax Estimates}
\label{sec:comp}

\subsection{Absolute Magnitude of RRL}

Since the work of Clube \& Dawe (1980a,b) there have been multiple attempts to derive the RRL absolute magnitude scale from \statpi\ (e.g. Hawley et al. (1986), Strugnell, Reid \& Murray (1986),  Layden et al. (1996), PG$^3$, Luri et al. (1998), Dambis (2009)).  The most recent \statpi\ estimate of $M_{V,RR}$\footnote{Dambis et al. (2009) computed the \statpi\ for a sample of 364 Galactic RRL stars in the Northern Hemisphere from targets in the Beers (2000) catalog using 2MASS photometry to correct the infrared-inferred period-luminosity relation for RRL stars, finding  $\langle M_{K_s}({\rm Adopted}) \rangle=-2.33\log(P_F) -0.818 \pm 0.081$.   This work does not estimate $M_{V,RR}$ and can thus not be directly compared here.} (and so directly comparable to the work  presented here) is by PG$^3$ who found 

\begin{equation}
M_V=0.75 \pm 0.13 \quad {\rm  at} \quad \langle {\rm [Fe/H]} \rangle =-1.6
\end{equation}
from a sample of 182 halo RRab stars taken from Layden et al. (1996)\footnote{Luri et al. (1998) found a similar value from a smaller sample (144 stars total) than PG$^3$.}

Benedict et al. (2011) used HST to obtain trigonometric parallaxes for 5 RRL (4 RRab and 1 RRc).  Combining their reported values and measurement errors (together with their adopted intrinsic dispersion of $\sigma_{M_V} = 0.0577$) yields 
\begin{equation}
M_V = 0.443 \pm 0.067\quad {\rm  at} \quad \langle {\rm [Fe/H]} \rangle =-1.50
\end{equation}
while, using $\sigma_{M_V} = 0.1$ (adopted here) yields $M_V=0.426\pm 0.080$
at $\langle {\rm [Fe/H]} \rangle =-1.52$.
To obtain these results, we weight the individual measured magnitude offsets
from the mean by $(\sigma_{M_V}^2 + K^2\sigma_{\rm [Fe/H]}^2)^{-1}$, where
$K=0.214$ is the slope of the $M_V$-[Fe/H] relation.  Note that Benedict et al. (2011) incorrectly quote somewhat smaller errors.  See  Appendix B.

Efforts to measure $M_V$ from RRc stars have been much more limited. Hawley et al. (1986) and Strugnell, Reid \& Murray (1986) obtained \statpi\ results from 17 and 26
candidate RRc stars respectively, but were forced to make a restricted analysis
because of small number statistics, which they reported only
for ``completeness".  In addition, as mentioned above, Benedict et al.
(2011) obtained $M_V=0.27 \pm 0.17$ for a single RRc, RZ Cep at [Fe/H]=-1.77.

\subsection{Velocity Ellipsoid}
Our results on the kinematics of the halo ($(W_\pi, W_\theta, W_z) = (10.9,34.9,7.2)\kms$ with dispersions $(\sigma_{W_\pi}, \sigma_{W_\theta}, \sigma_{W_z}) =  (154.7, 103.6, 93.8) \kms$) are in excellent agreement with previous work.  Comparing with PG$^3$, our results are consistent at well under 1-$\sigma$ despite an entirely different sample.  Comparing with the most recent (and largest) \statpi\ analysis of Dambis et al. (2009) our results are also generally in excellent agreement, as almost all parameters of the halo {\it and} disk velocity ellipsoids agree within 1 or 2-$\sigma$ with the exception of vertical velocity dispersions of the halo which are in tension at the 3-$\sigma$ level.  We do not know the origin of this discrepancy, but will have more leverage to investigate this with our larger RRab sample.

\section{Discussion and Conclusions}
We have performed the first decisive analysis of the absolute magnitude for RRc variables via statistical parallax using the first data from the Carnegie RR Lyrae Survey (CARRS).  Our current measurements for RRc variables yield a 5\% distance error which is similar to that obtained from modern techniques applied to RRab samples.  We find a velocity ellipsoid for our disk and halo population that is in good agreement with previous measurements.  

Already, CARRS provides competitive distance accuracy to other surveys and techniques.  At the conclusion of CARRS, we anticipate a factor of $\sim 4$ increase in the number of tracers and, consequently, 2\% distance errors.  In future work, we will analyze this far larger database and have the statistical potency to divide our sample into finer metallicity and kinematic bins than can be done presently.  This will allow a precision measurement for comparison with other techniques with the hope of a ``unified" RRL distance scale.  

In the Gaia era, where space-based parallaxes will be available for many of these objects, our database of high-resolution spectra should provide useful complementary information for going beyond distances and gaining further understanding of RRL as astrophysical objects, rather than merely as test particles.

\appendix
\begin{appendices}

\section{A: Analytic Estimates of \statpi\ Correlation Coefficients}

In the main text, we showed that the analytic error estimates 
derived by PG98a for the 10 \statpi\ parameters closely approximate
the numerical errors in our sample, despite the fact that the
analytic treatment assumes zero measurement errors, uniform sky
coverage, and no rotation with Galactic position.  Here, we extend
the PG98a analytic treatment to the off-diagonal elements of the
parameter determinations and compare the resulting correlation coefficients
to our numerical values. 

Integrating Equations (30)-(35) from PG98a over $N$ objects uniformly
distributed on the sky, we obtain the inverse covariance matrix of
the 10 \statpi\ parameters 
$a_0 \ldots a_9$ $=(\ln\eta,W_i,C_{ii},C_{i\not= j})$ 
\begin{equation}
b_{\nu 0} = b_{0 \nu} = N A_\nu;
\qquad
b_{mn} = N B_m \delta_{mn}
\end{equation}
where
\begin{equation}
A_0 = 4 + {2\kappa^2\over 3}.
\quad
A_{1-3} = -{2 W_i\over 3 C_{ii}},
\quad
A_{4-6} = -{4\over 3 C_{ii}},
\quad
A_{7-9} = 0,
\end{equation}
\begin{equation}
B_{1-3} = {1\over C_{ii}},
\quad
B_{4-6} = {2\over C_{ii}^{2}},
\quad
B_{7-9} = {1\over C_{ii}C_{jj}},
\end{equation}
and where we have adopted a reference frame in which $C_{ij}$ is diagonal.
(See Equations (37)-(42) of PG98a.)

Matrices of this form can be analytically inverted, $c\equiv b^{-1}$ as
\begin{equation}
c_{00} =\biggl(A_0 -\sum_m {A_m^2\over B_m}\biggr)^{-1},
\quad
c_{0m} = c_{m0} = -{A_m\over B_m}c_{00},
\quad
c_{mn} = {\delta_{mn}\over B_m}+{A_m A_n\over B_m B_n}c_{00}.
\quad
{\label{eqn:cov}}
\end{equation}
PG98a (Equations (43)-(46))
have already evaluated the diagonal elements of this matrix
(i.e., the variances of the parameters)
\begin{equation}
\var(\ln\eta) = {\alpha\over N}
\quad
{\var(W_i)\over C_{ii}}= {1\over N}\biggl(1 + {4\over 9}{W_i^2\over C_{ii}}
\alpha \biggr),
\quad
{\var(C_{ii})\over C_{ii}^2}= {1\over N}\biggl(1 + {8\over 9}\alpha \biggr),
\quad
{\var(C_{i\not=j})\over C_{ii}C_{jj}}= {1\over N},
\end{equation}
where $\alpha^{-1}\equiv 4/3 + (2/9)\kappa^2$.  Here we use
Equation(\ref{eqn:cov}) to evaluate the off-diagonal elements, or equivalently
the correlation coefficients
\begin{equation}
\cc(\ln\eta,W_i) = Q_{W,i}
\quad
\cc(\ln\eta,C_{ii}) = Q_{C,i}
\end{equation}

\begin{equation}
\cc(W_i,W_j) = Q_{W,i}Q_{W,j}
\quad
\cc(W_i,C_{jj}) = Q_{W,i}Q_{C,j}
\quad
\cc(C_{ii},C_{jj}) = Q_{C,i}Q_{C,j}
\end{equation}
where
\begin{equation}
Q_{W,i}\equiv {\rm sgn}(W_i)\biggl(1+ {9 C_{ii}\over 4 W_i^2\alpha}\biggr)^{-1/2},
\quad
Q_{C,i}\equiv\biggl(1+ {9\over 8\alpha}\biggr)^{-1/2},
\end{equation}
and all other terms vanish.  Note that in making these evaluations,
one must return to the ``Sun frame'', i.e. add $(9,250,7)\,\kms$ back
to the $W_i$.

In the matrix below, we compare the analytic correlation coefficients
(above diagonal) to the actual ones (below diagonal).  There is
good overall agreement.
$$
\left( \begin{array}{c c c c c c c c c c}
  1.00 & 0.01 &-0.68 & 0.01 & 0.53 & 0.53 & 0.53 & 0.00 & 0.00 & 0.00\\
 -0.02 & 1.00 &-0.01 & 0.00 & 0.00 & 0.00 & 0.00 & 0.00 & 0.00&  0.00\\
 -0.66 & 0.04 & 1.00 & 0.00 &-0.36 &-0.36& -0.36 & 0.00 & 0.00 & 0.00\\
 -0.04 & 0.00 &-0.04 & 1.00 & 0.00 & 0.00 & 0.00 & 0.00 & 0.00&  0.00\\
  0.60 &-0.01 &-0.40 &-0.03 & 1.00 & 0.28&  0.28 & 0.00 & 0.00&  0.00\\
  0.38 &-0.01 &-0.25 & 0.00 & 0.23 & 1.00 & 0.28 & 0.00 & 0.00 & 0.00\\
  0.41 & 0.00 &-0.27 &-0.02 & 0.25 & 0.15 & 1.00 & 0.00 & 0.00 & 0.00\\
 -0.16 & 0.01&  0.11 & 0.01 &-0.08 &-0.06 &-0.08 & 1.00 & 0.00 & 0.00\\
  0.00 &-0.01 &-0.01 & 0.00 & 0.00 & 0.01 &-0.01& -0.10 & 1.00 & 0.00\\
  0.08 & 0.00 &-0.06 &-0.02 & 0.06 & 0.00& -0.02 & 0.01 & 0.03 & 1.00\\
  \end{array}\right)
$$

We note that the errors in these parameters are almost perfectly 
Gaussian, and therefore completely described by their means and
covariances.  This can be seen as follows.  In any linear fit,
the parameter estimates, $a_i$ can be written as a linear function of
the data $y_l$: $a_i = c_{ij}d_j$, $d_j=\sum_{kl} {\cal B}_{kl}f_j(t_k) y_l$,
where $f_j(t_k)$ are the trial functions, ${\cal B}_{kl}$ is the inverse
covariance matrix of the data, $c\equiv b^{-1}$, and 
$b_{ij} = \sum_{kl}{\cal B}_{kl}f_i(t_k)f_j(t_l)$.  Therefore if the data
$y_l$ have Gaussian errors, then so do the parameters $a_i$.  Note that
this statement does not depend in any way on the central limit theorem,
as is sometimes supposed, but only on the fact that linear combinations
of Gaussians are Gaussians. 

In the present problem, $-2\ln L$ looks similar in structure to a 
standard linear-fit $\chi^2$
except, first, the parameters appear in non-linear combinations,
second, the second moments of the velocity ellipsoid appear in an
additional term containing a log determinant, and third, the
scaling factor $\eta$ also appears in a log term.  Nevertheless,
the linearized problem (treated explicitly by PG98a) is extremely
close in structure to a standard linear-fit $\chi^2$.  Therefore,
one expects similar mathematical properties.  This is the underlying
reason that the PG98a linearized analysis matches numerical
results so closely.

\section{B. Appropriate Weighting in Deriving Uncertainties}

Here we derive the appropriate weighting scheme for an ensemble
of RRL parallax measurements with different errors in both
$M_V$ and [Fe/H].

Let us consider $n$ stars with measured absolute magnitudes
$M_{V,i}$ and errors $\sigma_i$ and each with perfectly known
metallicity $\feh_i$.  And let us initially assume that the
slope of the $\feh-M_V$ relation is known but the zero-point
is not.  That is,

\begin{equation}
M_{V,\rm pred}(\feh) = a + K(\feh - Q)
\end{equation}
where $Q$ is some arbitrarily chosen fiducial metallicity,
$K$ is the known slope and $a$ is the unknown zero point.
Then
\begin{equation}
\chi^2 = \sum_{i=1}^n 
\biggl({M_{V,\rm pred}(\feh_i) - M_{V,i}\over \sigma_i}\biggr)^2
\end{equation}
This $\chi^2$ is minimized by setting its derivative to zero, i.e.,
\begin{equation}
a\sum_i w_i = \sum_i w_i (M_{V,i} - K(\feh_i - Q)),
\qquad w_i \equiv \sigma_i^{-2}
\end{equation}

We now choose the ``arbitrary'' fiducial metallicity so that the
relation is independent of the choice of $K$.  This can be done
if we choose
\begin{equation}\label{eqn:q}
Q = {\sum_i w_i \feh_i\over \sum_i w_i}
\end{equation}
in which case $\chi^2$ is minimized at
\begin{equation}
a = {\sum_i w_i M_{V_i}\over \sum_i w_i},
\end{equation}
regardless of $K$.
The error in this estimate is the point at which $\chi^2$ rises by
unity, i.e.,
\begin{equation}
\sigma(a) = \sqrt{2\over d^2\chi^2/d a^2} = \bigl(\sum_i w_i\bigr)^{-1/2}
\end{equation}
Hence, Equation(\ref{eqn:q}) gives the effective metallicity at which
the measurement is made.

Now let us suppose that the metallicities are given to us with error
bars $\Sigma_i$, so that we can consider simultaneously fitting for
both the zero-point (now called $a_0$) and the five true metallicities,
called $a_i$.  We can write $\chi^2$
\begin{equation}
\chi^2 = \sum_{i=1}^n 
\biggl({M_{V,\rm pred}(a_i) - M_{V,i}\over \sigma_i}\biggr)^2 + 
\biggl({a_i - \feh_i\over \Sigma_i}\biggr)^2 
\end{equation}
Setting the $n+1$ derivatives of $\chi^2$ to zero (wrt to $a_0,a_i$)
yields the equation
\begin{equation}
\sum_{\nu=0}^n b_{\mu\nu}a_\nu = d_\mu
\end{equation}
where
\begin{equation}
d_0 = \sum_{i=1}^n w_i(M_{V,i} + KQ),
\qquad
d_i = w_i (M_{V,i} + KQ)K + W_i\feh_i
\quad
W_i \equiv \Sigma_i^{-2}
\end{equation}
\begin{equation}
b_{00} = A_0
\qquad
b_{0i} = b_{i0}= A_i
\qquad
b_{ij} = \delta_{ij}B_i
\end{equation}
and
\begin{equation}
A_0 \equiv \sum_{i=1}^n w_i
\qquad
A_i\equiv K w_i,
\qquad
B_i\equiv K^2 w_i  + W_i
\end{equation}

The inverse of this matrix, $c\equiv b^{-1}$ is given by Equation(\ref{eqn:cov}), which then allows us to evaluate $a_0=\sum_\nu c_{0\nu} d_\nu$,
\begin{equation}
a_0 = c_{00}\biggl[\sum_{i=1}^n w_i(M_{V,i} + KQ)
\biggl(1 - {b_{0i}\over b_{ii}}K\biggr) - 
\sum_{i=1}^n W_i\feh_i{b_{0i}\over b_{ii}}\biggr]
\end{equation}
This simplfies to
\begin{equation}
a_0 = c_{00}\sum_{i=1}^n
{M_{V,i} - K(\feh_i - Q)\over
(K\Sigma_i)^2 + \sigma_i^2};
\qquad
c_{00}^{-1} = \sum_{i=1}^n
{1\over (K\Sigma_i)^2 + \sigma_i^2};
\end{equation}
where we note that $c_{00}^{1/2}$ is the error in $a_0$ (e.g., Gould 2003a). This looks identical to our previous expression that ignored the $\feh$ errors, except the inverse weights are increased
fractionally by $(K\Sigma/\sigma)^2$.
This result formally confirms one's naive idea that the metallicity
error is ``equivalent'' to an additional error in the absolute
magnitude, propagated by the slope $(K)$ of the relation.

Note, however, that in contrast to the case of perfectly known metallicities,
one cannot enforce complete independence of the result from choice of
$K$ simply by adopting $Q$ as the average metallicity weighted by
$[(K\Sigma)^2 + \sigma^2]^{-1}$, since the weighting itself depends
on $K$. However, in the present case $(K\Sigma/\sigma)^2\ll 1$,
so this has no practical impact.

\end{appendices}

\clearpage
\end{document}